\documentclass[aps,prb,twocolumn,floatfix,showpacs,groupedaddress]{revtex4-2}
\usepackage{graphicx,amsfonts,bbold,bm}
\usepackage{color}

\begin{document}
\title{Linear Response Based Theories for Dzyaloshinskii-Moriya Interactions}
\author{I.~V.~Solovyev}
\email{SOLOVYEV.Igor@nims.go.jp}
\affiliation{National Institute for Materials Science, MANA, 1-1 Namiki, Tsukuba, Ibaraki 305-0044, Japan}
\date{\today}

\begin{abstract}
We investigate abilities of various linear response based techniques for extracting parameters of antisymmetric Dzyaloshinskii-Moriya (DM) interactions from the first-principles electronic structure calculations. For these purposes, we further elaborate the idea of Sandratskii [Phys. Rev. B \textbf{96}, 024450 (2017)], which states that the $z$ component of the DM vector can be computed by retaining only the \emph{spin-diagonal} part of the spin-orbit (SO) interaction. This approximation, which becomes exact to the first order in the SO coupling, greatly simplifies the calculations as it requires only minor extensions in comparison to isotropic parameters in the nonrelativistic case. We start our analysis with the magnetic force theorem (MFT), which relies on additional approximations resulting in the linear dependence of the exchange interactions on the response tensor, and compare it with the exact approach formulated in terms of the inverse response. For the ligand states, which are not primarily responsible for the magnetism but magnetised from the localized states, we propose the downfolding procedure transferring the effect of these ligand spins into parameters of effective interactions between the localized spins. These techniques are applied for the series of CrCl$_3$ and CrI$_3$ based materials, including bulk, monolayer, bilayer, and three-layer systems. Particularly, we discuss how the DM interactions are induced by the inversion symmetry breaking at the surface or by the electric field. As long as the response tensor between the Cr $3d$ states is calculated by taking into account the SO interaction on the heavy ligand sites, the MFT appears to be a good approximation for the DM interactions, being in contrast with the isotropic exchange, for which MFT and the exact method provide quite a different description. Finally, we discuss the relevance of our approach to other techniques ever proposed for calculations of the DM interactions. Particularly, we argue that the spin-current model for the DM interactions can be derived from the MFT based expression and is the relativistic counterpart of the double exchange, occurring in metallic systems in the limit of infinite exchange splitting.
\end{abstract}
\maketitle

\section{\label{sec:Intro} Introduction}
\par The Dzyaloshinskii-Moriya (DM) interaction is the very peculiar type of the exchange coupling emerging between noncollinear spins. The existence of such interaction was predicted by Dzyaloshinskii in 1958, on the basis of symmetry arguments~\cite{Dzyaloshinskii_weakF}. The microscopic picture behind this interaction was proposed by Moriya two years later, as an effect of relativistic spin-orbit (SO) coupling in noncentrosymmetric bonds~\cite{Moriya_weakF}. Competing with the isotropic exchange, the DM interaction results in deformation of the conventional ferromagnetic or antiferromagnetic order and the formation of noncollinear magnetic textures, including the spin canting~\cite{Dzyaloshinskii_weakF}, spin spirals~\cite{Dzyaloshinskii_helix}, magnetic skyrmions~\cite{bog1,bog2}, etc. The corresponding spin model, describing this competition, is given by
\noindent
\begin{equation}
{\cal E} =  - \frac{1}{2} \sum_{i \ne j} \big( J_{ij} \boldsymbol{e}_{i} \cdot \boldsymbol{e}_{j} - \boldsymbol{d}_{ij} \cdot [\boldsymbol{e}_{i} \times \boldsymbol{e}_{j}] \big),
\label{eq:Hspin}
\end{equation}
\noindent where $J_{ij}$ is the isotropic exchange, $\boldsymbol{d}_{ij}$ is the DM vector, and $\boldsymbol{e}_{i}$ is the unit vector in the direction of spin at the site $i$.

\par Recently, the growing interest in the physics of DM interaction has been spurred by the discoveries of new phenomena, such as the topological Hall effect~\cite{topologicalHall}, chiral domain walls~\cite{DomainWalls1,DomainWalls2}, the magnetoelecric coupling in spiral magnets~\cite{Kimura_TbMnO3,KNB,1orbital}, etc., which are intrinsically related to the noncollinear alignment of spins.

\par In order to clarify the microscopic mechanisms responsible for these phenomena and systematically search for the new materials, where these phenomena can be realized, there is strong demand for theoretical electronic structure calculations based on the density functional theory (DFT)~\cite{HK}. One of the main tasks here is to develop numerical methods for realistic calculations of the DM interactions. However, the situation in this area appears to be rather diverse, as there is already quite a few number of such methods, which rely on different assumptions and approximations. (i) The DM interactions can be evaluated in the framework of the superexchange theory~\cite{Yildirim}, following the original idea by Moriya~\cite{Moriya_weakF}, and extracting the parameters of such theory from the first-principles calculations~\cite{1orbital,PRB2015b}. However, the superexchange is the theory of Mott insulators, which is valid in the limit of large Coulomb repulsion on the atomic sites and does not apply for metallic systems~\cite{Anderson}. (ii) Another possibility is to define the spin model (\ref{eq:Hspin}) locally, by considering the infinitesimal rotations of spins near certain magnetic equilibrium and employing for these purposes the magnetic force theorem (MFT), as it was originally proposed by Liechtenstein \textit{et al.}~\cite{LKAG1987}. In this case, MFT means the perturbation theory for the energy change formulated in terms of the real space Green's functions. Alternatively, it can be reformulated in terms of the response function (or transverse spin susceptibility) in the reciprocal space~\cite{KL2000}. There is a number of relativistic extensions of this method dealing with the full Green function, which incorporates all possible effects of the SO interactions and takes into account the coupling between states with the same as well as opposite projections of spin~\cite{AKL1997,Ebert,Kvashnin2020,Mahfouzi}. For each bond, such technique allows us to derive the $3$$\times$$3$ exchange tensor, which can be further decomposed into the isotropic exchange, antisymmetric DM interaction, and symmetric exchange anisotropy. (iii) Other techniques, which are conceptionally similar to MFT~\cite{LKAG1987}, include mixed perturbation theory with respect to the infinitesimal rotations of spins and the SO interactions~\cite{PRL96}, spin susceptibility~\cite{Koretsune} and Berry phase theory~\cite{Freimuth} methods. (iv) On the other hand, Sandratskii has argued that in order to calculate $z$ component of the DM vector, $d^{z}_{ij}$, it is sufficient to know only the change of the electronic structure caused by the SO interaction separately for the majority- and minority-spin states, and neglecting the coupling between opposite projections of spin~\cite{Sandratskii}. Obviously, this is the very strong statement, which greatly simplifies the calculations and allows us to use for these purposes the generalized Bloch theorem~\cite{Sandratskii,Sandratskii_review}. (v) The validity of the MFT itself, at least in the way how it was formulated in Ref.~\cite{AKL1997}, was questioned~\cite{BrunoPRL2003,Streib,PRB2021}. The key problems are the missing constraining field, which is required in order to fix the directions of spins, and how to deal with the ligand spins~\cite{PRB2021,Arxiv2022}. (vi) The spin-current method~\cite{Katsnelson_DM,Kikuchi} typically stays separately from the above mentioned techniques. For instance, its relationship with MFT is unknown and the DM parameters do not explicitly depend on the eenrgy splitting between the majority and minority spins as it would be expected from the model analysis, at least for insulating systems~\cite{Moriya_weakF,Yildirim}.

\par The main purpose of this work is to rationalize these methods by presenting an unified point of view on how the DM interactions can be generally calculated on the basis of DFT. We start with the idea proposed by Sandratskii~\cite{Sandratskii} and show how it can be used within MFT framework developed by Liechtenstein \textit{et al.}~\cite{LKAG1987} (Sec.~\ref{sec:MFT}). It provides a transparent expression for $d^{z}_{ij}$, which can be computed at the same cost as the isotropic exchange. Then, we switch to the exact formalism (Sec.~\ref{sec:exact}) and show how the ligand spins can be rigorously eliminated so that their contributions become included to the effective DM interactions between the localized spins (Sec.~\ref{sec:downfolding}). The abilities of these methods and the main tendencies in the behavior of DM interactions are illustrated for CrCl$_3$ and CrI$_3$, in their bulk, monolayer, bilayer, and three-layer realization (Sec.~\ref{sec:Results}). The alternative techniques, including mixed perturbation theory~\cite{PRL96} and spin-current model~\cite{Katsnelson_DM,Kikuchi}, are discussed in Sec.~\ref{sec:Alternative}. Particularly, we show how the spin-current model can be derived from the MFT based expression, pretty much similar to the canonical double exchange in the nonrelativistic case~\cite{DE,PRL1999}. Finally, a brief summary will be given in Sec.~\ref{sec:summary}.

\section{\label{sec:Method} Theory}
\par Our starting point for the analysis of interatomic exchange interactions is a spin-dependent one-electron tight-binding Hamiltonian $\hat{H}^{\sigma}$ (where $\sigma$$=$ $\uparrow$ or $\downarrow$), formulated in the basis of transition-metal $d$ and ligand $p$ states. In practice, such Hamiltonian can be rigorously derived from the spin-polarized DFT~\cite{HK}, by solving the Kohn-Sham (KS) problem~\cite{KS,BarthHedin} and recalculating the Hamiltonian matrix in the appropriate Wannier basis~\cite{WannierRevModPhys,JPCMreview}. The key point here is that the knowledge of such one-electron Hamiltonian is sufficient for describing the energy change caused by the infinitesimal rotations of spins, as was advocated by MFT~\cite{AFT,LKAG1987}.

\subsection{\label{sec:MFT} Magnetic force theorem}
\par In order to illustrate the basic idea of our method, it is convenient to start with the particular realization of MFT formulated by Liechtenstein \textit{et al.} in Ref.~\cite{LKAG1987}, which assumes that (i) the rotation of the magnetization, $\hat{m}^{z}_{i} \to (\sin \theta, 0, \cos \theta )\hat{m}^{z}_{i}$, produces the rotation of the exchange-correlation (xc) field by the same angle, $\hat{b}^{z}_{i} \to (\sin \theta, 0, \cos \theta )\hat{b}^{z}_{i}$, and (ii) this is the only perturbation caused by the infinitesimal rotations of spins by the given angle. While the first assumption can be generally proved~\cite{PRB1998}, the second one is an approximation, as it ignores another important contribution -- the external field, constraining the direction of the spin magnetization~\cite{BrunoPRL2003,Streib,PRB2021}. We will continue to name this  approximate scheme as ``MFT'', though it should be understood that the ``exact approach'', which we propose as the alternative to it, is nothing but the rigorous realization of MFT, which allows us to relate the energy change caused by the infinitesimal rotations of spins with the properties of $\hat{H}^{\sigma}$ in the ground state.

\par
\begin{figure}[t]
\begin{center}
\includegraphics[width=8.6cm]{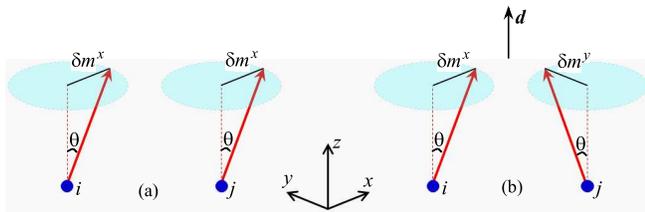}
\end{center}
\caption{Type of infinitesimal spin rotations, which can be used in the Green function perturbation theory in order to calculate the isotropic exchange (a) and Dzyaloshinskii-Moriya interaction $\boldsymbol{d}$ along the $z$ axis (b).}
\label{fig.cartoon}
\end{figure}

 The basic idea of Liechtenstein \textit{et al.} is to treat the change of the xc field as a perturbation in the framework of the Green's function method~\cite{LKAG1987}. For instance, considering the perturbation caused by the $x$ components of the xc field at the sites $i$ and $j$ without the SO interaction (see Fig.~\ref{fig.cartoon}a), it is straightforward to derive the following expression for the isotropic exchange~\cite{LKAG1987}:
\noindent
\begin{widetext}
\begin{equation}
J_{ij} = \frac{1}{4 \pi} {\rm Im} \int_{- \infty}^{\varepsilon_F} d \varepsilon \, {\rm Tr}_{L} \left\{ \hat{b}^{z}_{i} \hat{G}^{\uparrow}_{ij}(\varepsilon) \hat{b}^{z}_{j} \hat{G}^{\downarrow}_{ji}(\varepsilon) + \hat{b}^{z}_{i} \hat{G}^{\downarrow}_{ij}(\varepsilon) \hat{b}^{z}_{j} \hat{G}^{\uparrow}_{ji}(\varepsilon) \right\},
\label{eq:JijMFT}
\end{equation}
\end{widetext}
\noindent where $\hat{G}_{\phantom{ij}}^{\sigma} = \big(  \varepsilon - \hat{H}^{\sigma} \big)^{-1}$, ${\rm Tr}_{L}$ is the trace over the orbital indices, and $\hat{b}^{z}_{i}$ is associated with the site diagonal part of $\hat{H}^{\sigma}$~\cite{PRB2021}. In this formulation, $J_{ij}$ can be viewed as the interaction between the fluctuations $\delta \hat{b}^{x}_{i} = \theta \hat{b}^{z}_{i}$ and $\delta \hat{b}^{x}_{j} = \theta \hat{b}^{z}_{j}$ along the same axis $x$. Here, the corresponding energy change, formulated via the Green functions, is associated with the interaction $-J_{ij}\delta m^{x}_{i} \delta m^{x}_{j}$ between $m^{x}_{i} \sim \theta$ and $m^{x}_{j} \sim \theta$ in the Heisenberg model, which yields Eq.~(\ref{eq:JijMFT}) for the isotropic exchange~\cite{LKAG1987}.
\noindent

\par Then, following the same arguments, one can consider the interaction between the perpendicular fluctuations along $x$ and $y$, $\delta \hat{b}^{x}_{i}$ and $\delta \hat{b}^{y}_{i}$, which should give us the $z$ component of the DM vector (see Fig.~\ref{fig.cartoon}b). Without SO coupling, we would have the following expression, where the energy change is mapped onto $d^{z}_{ij}\delta m^{x}_{i} \delta m^{y}_{j}$:
\noindent
\begin{widetext}
\begin{equation}
d_{ij}^{z} = \frac{1}{4 \pi} {\rm Im} \, i \int_{- \infty}^{\varepsilon_F} d \varepsilon \, {\rm Tr}_{L} \left\{ \hat{b}^{z}_{i} \hat{G}^{\uparrow}_{ij}(\varepsilon) \hat{b}^{z}_{j} \hat{G}^{\downarrow}_{ji}(\varepsilon) - \hat{b}^{z}_{i} \hat{G}^{\downarrow}_{ij}(\varepsilon) \hat{b}^{z}_{j} \hat{G}^{\uparrow}_{ji}(\varepsilon) \right\}.
\label{eq:dijzMFT}
\end{equation}
\end{widetext}
\noindent Obviously, if the sites $i$ and $j$ are connected by the spacial inversion, the two terms in $\{ \ldots \}$ are equivalent and, therefore, $d_{ij}^{z} = 0$, as it should be for the DM interactions~\cite{Dzyaloshinskii_weakF,Moriya_weakF}. Moreover, there are additional symmetries, which enforce $d_{ij}^{z} = 0$ without the SO coupling. Namely, since the time reversal symmetry is broken only in the spin sector, the one-electron Hamiltonian is invariant under the complex conjugation, $( \hat{H}^{\sigma} )^{*} = \hat{H}^{\sigma}$, so as the Green function, leading to the property $G_{ij}^{ab} = \big( G_{ij}^{ab} \big)^{*} = G_{ji}^{ba}$, which holds in the basis of real harmonics ($a$ and $b$) for each projection of spin ($\uparrow$ or $\downarrow$). Furthermore, $\hat{b}^{z}_{i}$ is the real symmetric matrix. Therefore, the two terms in $\{ \ldots \}$ remain equivalent yielding $d_{ij}^{z} = 0$.

\par The SO interaction leads to the coupling between the majority ($\uparrow$) and minority ($\downarrow$) spin states. Nevertheless, if our goal is just to calculate $d_{ij}^{z}$, we do not need to consider such coupling. This important point was realized by Sandratskii~\cite{Sandratskii}. His arguments are based on the observation that if the spins form the spiral structure in the $xy$ plane, $\boldsymbol{e}_{i} = (\cos \boldsymbol{q} \cdot \boldsymbol{R}_{i}, \sin \boldsymbol{q} \cdot \boldsymbol{R}_{i}, 0)$ (and considering here, for simplicity, the case where there is only one magnetic site in the primitive cell), the energy of the spin model (\ref{eq:Hspin}) is uniquely specified by the spin-spiral propagation vector $\boldsymbol{q}$:
\noindent
\begin{equation}
{\cal E}(\boldsymbol{q}) = -\frac{1}{2} \big\{ J(\boldsymbol{q}) -i d^{z}(\boldsymbol{q}) \big\},
\label{eq:Hspinq}
\end{equation}
\noindent where $X(\boldsymbol{q}) = \sum_{i} X_{0i} \exp(-i \boldsymbol{q} \cdot \boldsymbol{R}_{i})$ for $X=$ $J$ or $d^{z}$ and $\boldsymbol{R}_{i}$ denotes the position of the site $i$. If the spin model is obtained by mapping the total energies of a more general electronic model, these energies should have a similar form and be uniquely specified by $\boldsymbol{q}$. This can be achieved by employing the generalized Bloch theorem, which allows us combine the periodic translations with the $SU(2)$ rotations of electronic states in the spin subspace and, for each $\boldsymbol{q}$, reduce the computational efforts to the solution of the problem within just one primitive cell, as in the collinear case~\cite{Sandratskii_review}. This should give us ${\cal E}(\boldsymbol{q})$, which can be used for the mapping onto Eq.~(\ref{eq:Hspinq}). However, the generalized Bloch theorem is no longer applicable when the SO interaction is taken into account. The reason is that the spin-off-diagonal part of the SO interaction is not invariant under the $SU(2)$ rotations. The basic idea of Sandratskii in this respect is that, since there should be one to one correspondence between the energies of electronic and spin models, one should keep only spin-diagonal SO interaction, which remains invariant under the $SU(2)$ rotations and, thus, allows us to use the generalized Bloch theorem~\cite{Sandratskii}.

\par This can be paraphrased differently. The SO coupling gives rise to the antisymmetric DM interactions as well as the magnetic anisotropy, which act differently on the magnetic alignment. While the DM interactions favor the spin-spiral order, the magnetic anisotropy typically acts against it, by deforming the spin spiral~\cite{Koehler} and tending to lock the propagation vector $\boldsymbol{q}$ to commensurate values~\cite{PRB2011}. Therefore, one should find a way how to separate the contributions to the DM interactions and magnetic anisotropy in the electronic structure calculations. One possibility, if we want to calculate the DM interactions, is to restore the validity of the generalized Bloch theorem. However, the price to pay is to neglect the off-diagonal elements of the SO interaction, which contribute to the magnetic anisotropy.

\par Another intuitive reason why for $d^{z}$ it is sufficient to consider only spin-diagonal elements of the SO coupling is that the DM interactions are related to the energy change emerging in the 1st order of the SO coupling. Therefore, if we start with the collinear configuration of spins, this energy change will include only diagonal elements of the SO coupling.

\par In what follow, $d_{ij}^{z}$ can still be calculated using Eq.~(\ref{eq:dijzMFT}), but with the Green function $\hat{\cal G}_{\phantom{ij}}^{\sigma} = \big(  \varepsilon - \hat{H}^{\sigma} \mp \frac{1}{2}\xi \hat{L}^{z} \big)^{-1}$, which would include the spin-diagonal SO interaction, $\pm \frac{1}{2}\xi_{i} \hat{L}^{z}$, where $\xi_{i}$ is the SO coupling parameter, $\hat{L}^{z}$ is the angular momentum, and the upper (lower) sign stands for $\sigma$$=$ $\uparrow$ ($\downarrow$). This $\pm \frac{1}{2}\xi_{i} \hat{L}^{z}$ acts as an effective magnetic field, which breaks the time reversal symmetry in the orbital sector, yielding finite $d_{ij}^{z}$ when it is allowed by the crystallographic symmetry. Two other components of the DM vector, $d_{ij}^{x}$ and $d_{ij}^{y}$, can be computed by rotating the coordinate frame and applying the same Eq.~(\ref{eq:dijzMFT}). Furthermore, the spin-diagonal part of the SO interaction should be included to calculate $d_{ij}^{z}$ but not $J_{ij}$. Otherwise, $J_{ij}$ will suffer from parasitic anisotropic exchange interactions, which are incomplete without the contributions arising from the spin-off-diagonal part of the SO coupling.

\par As long as we deal with the Green function in the real space, in the spirit of the MFT scheme by Liechtenshtein \textit{at al.}~\cite{LKAG1987}, it is not really necessary to keep only site-diagonal part of the SO coupling and omit the off-diagonal one. Instead, one can calculate the full Green function, including all possible contributions of the SO coupling, and evaluate the $3 \times 3$ exchange tensor by considering the infinitesimal rotations of the xc fields in each bond. In addition to the isotropic and DM interactions, such tensor will also include the symmetric anisotropic interactions~\cite{AKL1997,Ebert,Kvashnin2020,Mahfouzi}. The Green function based approach for the exchange interactions can be also generalized to include the local correlation effects by replacing the static KS potential and the xc field with the frequency dependent self-energy~\cite{KL2000}. Nevertheless, in order to calculate the DM interactions in the reciprocal space, by using the response tensor, it is crucial to bring in force the generalized Bloch theorem by retaining only spin-diagonal elements of the SO coupling. Particularly, in the next section, we will discuss the exact approach, which is formulated solely in terms of the response tensor, where such restrictions become inevitable. In principles, besides on-site elements, $\pm \frac{1}{2}\xi_{i} \hat{L}^{z}$, one can also consider inter-site elements of the SO coupling. The latter can play an important role in the model considerations~\cite{Katsnelson_DM}, where the contributions of the SO coupling are effectively included to the hopping parameters. Nevertheless, in order to use the generalized Bloch theorem, these matrix elements must be spin-diagonal.

\par
\begin{figure}[t]
\begin{center}
\includegraphics[width=8.6cm]{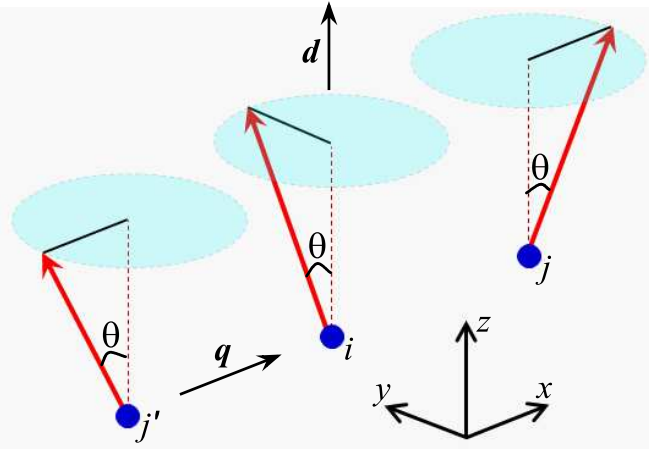}
\end{center}
\caption{Conical spin spiral, which is used for calculations of isotropic and Dzyaloshinskii-Moriya interactions in the linear response theory.}
\label{fig.cartoon3}
\end{figure}

 The response tensor in the reciprocal ($\boldsymbol{q})$ space is given by~\cite{KL2000}:
\noindent
\begin{widetext}
\begin{equation}
{\cal R}_{ab,cd}^{\sigma \sigma'} (\boldsymbol{q}) = \sum_{ml \boldsymbol{k}} \frac{f_{m \boldsymbol{k}}^{\sigma} - f_{l \boldsymbol{k}+\boldsymbol{q}}^{\sigma'}}{\varepsilon_{m \boldsymbol{k}}^{\sigma} - \varepsilon_{l \boldsymbol{k}+\boldsymbol{q}}^{\sigma'}} (C_{m \boldsymbol{k}}^{a\sigma})^{*}C_{l \boldsymbol{k}+\boldsymbol{q}}^{b\sigma'} (C_{l \boldsymbol{k}+\boldsymbol{q}}^{c\sigma'})^{*}C_{m \boldsymbol{k}}^{d\sigma},
\label{eq:gresponse}
\end{equation}
\end{widetext}
\noindent where $\sigma'$$=$ $\downarrow$ ($\uparrow$) if $\sigma$$=$ $\uparrow$ ($\downarrow$), $\varepsilon_{m \boldsymbol{k}}^{\sigma}$ are the KS eigenvalues, $| C_{l \boldsymbol{k}}^{\sigma} \rangle \equiv [ C_{l \boldsymbol{k}}^{a\sigma}]$ denotes the right (column) eigenvectors of tight-binding Hamiltonian, and $f_{m \boldsymbol{k}}^{\sigma}$ is the Fermi distribution function. The orbital indices in each of the pairs $ab$ and $cd$ belong to the same atomic sites in the unit cell. Instead of calculating the energies of pair interactions, the exchange parameters can be obtained by dealing with the energies of conical spin spirals with the propagation vectors $\boldsymbol{q}$ (see Fig.~\ref{fig.cartoon3}), which are mapped onto the spin model~(\ref{eq:Hspinq}) with several sites in the primitive cell~\cite{PRB2021}. More specifically, the $\theta$-dependence is treated analytically and compared with that of the spin model, describing the interactions between the transversal components of spins. Then, $J(\boldsymbol{q})$ and $d^{z}(\boldsymbol{q})$ are derived from the $\boldsymbol{q}$-dependence, following Eq.~(\ref{eq:Hspinq}) for the transversal spins.
\noindent
\noindent Thus, the isotropic exchange and DM interactions in the reciprocal $\boldsymbol{q}$-space can be calculated as
\noindent
\begin{equation}
J_{\mu \nu}(\boldsymbol{q}) = - \frac{1}{2} \big\{ \hat{b}^{z}_{\mu}  \mathcal{R}_{\mu \nu}^{+}(\boldsymbol{q})  \, \hat{b}^{z}_{\nu} - \hat{b}^{z}_{\mu} \hat{m}^{z}_{\nu} \delta_{\mu \nu} \big\}
\label{eq:jmft}
\end{equation}
\noindent and
\noindent
\begin{equation}
d_{\mu \nu}^{z}(\boldsymbol{q}) =  \frac{i}{2} \hat{b}^{z}_{\mu} \mathcal{R}_{\mu \nu}^{-}(\boldsymbol{q})  \, \hat{b}^{z}_{\nu} ,
\label{eq:dmft}
\end{equation}
\noindent respectively, where $\mathcal{R}_{\mu \nu}^{\pm}(\boldsymbol{q}) = \frac{1}{2} \big\{ \mathcal{R}_{\mu \nu}^{\uparrow \downarrow}(\boldsymbol{q}) \pm  \mathcal{R}_{\mu \nu}^{\downarrow \uparrow}(\boldsymbol{q}) \big\}$. Here, $\mathcal{R}_{\mu \nu}^{\sigma \sigma'}(\boldsymbol{q})$ is the shorthand notation for ${\cal R}_{ab,cd}^{\sigma \sigma'} (\boldsymbol{q})$, where the orbital indices $a$ and $b$ ($c$ and $d$) belong to the site $\mu$ ($\nu$) of the unit cell. In the MFT method, the exchange field $b^{z}_{\nu}$ and the magnetization density $m^{z}_{\nu}$ are typically treated as the matrices in the subspace of orbital indices and the construction $\hat{b}^{z}_{\mu} \mathcal{R}_{\mu \nu}^{\pm}(\boldsymbol{q})  \, \hat{b}^{z}_{\nu}$ implies the summation over these orbital indices of the repeated atomic sites. Up to this point, Eqs.~(\ref{eq:jmft}) and (\ref{eq:dmft}) are totally equivalent to, respectively, Eqs.~(\ref{eq:JijMFT}) and (\ref{eq:dijzMFT}), which are related to each other by the Fourier transform.

\par As discussed above, $\mathcal{R}_{\mu \nu}^{+}(\boldsymbol{q})$ should be computed without the SO interaction and only $\mathcal{R}_{\mu \nu}^{-}(\boldsymbol{q})$ should include the spin-diagonal part of the SO interaction. We would like to emphasize that the form of the response tensor (\ref{eq:gresponse}), where the Bloch states with the momentum $\boldsymbol{k}$ are coupled only to the states with the momentum $\boldsymbol{k}$$+$$\boldsymbol{q}$, implies that the KS eigenstates can be specified by the spin indices $\sigma = \uparrow$ or $\downarrow$. If it is not the case, and the indices $\sigma = \uparrow$ and $\downarrow$ are mixed by the SO interaction, the tensor is no longer represented in such form and will generally include the coupling between $\boldsymbol{k}$, $\boldsymbol{k}$$+$$\boldsymbol{q}$, as well as $\boldsymbol{k}$$-$$\boldsymbol{q}$~\cite{footnote}. Thus, although it is not spell out explicitly, the use of the spin susceptibility in the form of Eq.~(\ref{eq:gresponse}) for calculating the DM interactions~\cite{Freimuth,Koretsune} implies that the SO coupling is diagonal with respect to the spin indices and in this sense is totally quivalent to the approximation considered by Sandratskii~\cite{Sandratskii}.

\par Then, it is straightforward to show that
\begin{equation}
{\cal R}_{ab,cd}^{\downarrow \uparrow} (\boldsymbol{q}) = \big[ {\cal R}_{ba,dc}^{\uparrow \downarrow} (-\boldsymbol{q}) \big]^{*}.
\label{eq:Rproperty}
\end{equation}
\noindent Therefore, it is sufficient to calculate only ${\cal R}_{ab,cd}^{\uparrow \downarrow} (\boldsymbol{q})$ in full Brillouin zone, while ${\cal R}_{ab,cd}^{\downarrow \uparrow} (\boldsymbol{q})$ can be obtained from it using Eq.~(\ref{eq:Rproperty}). For these purposes, it is convenient to use the gauge-invariant tight-binding Hamiltonian: $\hat{H}^{\uparrow, \downarrow}(\boldsymbol{k}+\bm{G}) = \hat{H}^{\uparrow, \downarrow}(\boldsymbol{k})$, where $\bf{G}$ is the reciprocal translation. If the time-reversal symmetry is preserved in the orbital sector (e.g., without SO interaction), the Eq.~(\ref{eq:Rproperty}) becomes: ${\cal R}_{ab,cd}^{\downarrow \uparrow} (\boldsymbol{q}) = {\cal R}_{ba,dc}^{\uparrow \downarrow} (\boldsymbol{q})$.

\subsection{\label{sec:exact} Exact approach}
\par The exact expression for the energy change caused by the infinitesimal rotations of spins within spin-polarized DFT can be obtained by considering explicitly the constraining fields, which should be applied to the system in order to rotate the magnetization by the angle $\theta$ (see Figs.~\ref{fig.cartoon} and \ref{fig.cartoon3})~\cite{Streib,PRB2021}. This constraining field can be expressed via transversal fluctuations, $\delta m^{x}$ and $\delta m^{y}$, using the inverse response tensor~\cite{PRB2021}, yielding the following expression:
\noindent
\begin{equation}
J_{\mu \nu}(\boldsymbol{q}) = \frac{1}{2} \left( M_{\mu}^{z} {\mathbb Q}^{+}_{\mu \nu}(\boldsymbol{q}) M_{\nu}^{z} - B^{z}_{\mu} M_{\mu}^{z}\delta_{\mu \nu} \right),
\label{eq:jexactM}
\end{equation}
\noindent and
\noindent
\begin{equation}
d^{z}_{\mu \nu}(\boldsymbol{q}) = -\frac{i}{2} M_{\mu}^{z} {\mathbb Q}^{-}_{\mu \nu}(\boldsymbol{q}) M_{\nu}^{z},
\label{eq:dexactM}
\end{equation}
\noindent where ${\mathbb Q}^{\pm}_{\mu \nu}(\boldsymbol{q}) = \frac{1}{2} \big\{ \big[ {\mathbb R}^{\uparrow \downarrow} \big]^{-1} \pm \big[ {\mathbb R}^{\downarrow \uparrow} \big]^{-1}  \big\}_{\mu \nu}$,
${\mathbb R}_{\mu \nu}^{\sigma \sigma'}(\boldsymbol{q}) = \displaystyle\sum_{a \in \mu} \sum_{c \in \nu} {\cal R}_{aa,cc}^{\sigma \sigma'} (\boldsymbol{q})$, and $B_{\mu}^{z} = \frac{1}{n_{\mu}}{\rm Tr}_{L} \{ \hat{b}_{\mu}^{z} \}$ (with $n_{\mu}$ being the number of orbitals at the site $\mu$). The form of Eqs. (\ref{eq:jexactM}) and (\ref{eq:dexactM}) implies that the rotated object is the local spin moment $M_{\mu}^{z} = {\rm Tr}_{L} \{ \hat{m}_{\mu}^{z} \}$ rather than the magnetization density $\hat{m}_{\mu}^{z}$. The rotations of $M_{\mu}^{z}$ are less costly energetically and, therefore, more suitable to describe the low-energy excitations~\cite{PRB2021}.

\subsection{\label{sec:downfolding} Downfolding into the model of localized spins}
\par The basic idea of downfolding is to eliminate the spin degrees of freedom, which are not primarily responsible for the magnetism. For instance, the $3d$ spins of the transition-metal (${\rm T}$) sites typically participate as the source of the magnetism and are solely responsible for the spontaneous time-reversal symmetry breaking, while the spins of the ligand (${\rm L}$) sites basically follow the magnetic structure of the ${\rm T}$ sites via the hybridization effects. Thus, although the ${\rm L}$ sites are magnetized, this is a ``secondary effect'', which is induced by the magnetization of the ${\rm T}$ sites. Therefore, one can try to eliminate the spin degrees of freedom associated with the ${\rm L}$ states.

\par Without SO interaction, the total energy change caused by the infinitesimal rotations of spins will contain the contributions of the following types~\cite{PRB2021}:
\noindent
\begin{widetext}
\begin{equation}
\delta {\cal E} = - \frac{1}{2} \left( \, \theta_{\rm T}^{T} J_{\rm TT}^{\phantom{T}} \theta_{\rm T}^{\phantom{T}} + \theta_{\rm T}^{T} J_{\rm TL}^{\phantom{T}} \theta_{\rm L}^{\phantom{T}} + \theta_{\rm L}^{T} J_{\rm LT}^{\phantom{T}} \theta_{\rm T}^{\phantom{T}} + \theta_{\rm L}^{T} J_{\rm LL}^{\phantom{T}} \theta_{\rm L}^{\phantom{T}} \right),
\label{eq:eLT}
\end{equation}
\end{widetext}
\noindent which describes the interactions in the system of ${\rm T}$ and ${\rm L}$  spins as well as between them. Here, $J_{\rm AB}$ are the matrices in the subspace spanned by the atomic sites ${\rm A}$ and ${\rm B}$, $\theta_{\rm A}^{\phantom{T}}$ is the column vector of the polar angles specifying the rotations (see Fig.~\ref{fig.cartoon3}), and $\theta_{\rm A}^{T}$ is corresponding to it row vector. Eq.~(\ref{eq:eLT}) holds for each $\boldsymbol{q}$, which is dropped for simplicity. Then, for each instantaneous configuration of the ${\rm T}$ spins, the ${\rm L}$ coordinates can be found from the equilibrium condition $\frac{\partial} {\partial \theta_{\rm L}^{T}} \delta {\cal E} = 0$, which yields $\theta_{\rm L}^{\phantom{T}} = D_{\rm LT}^{\phantom{T}} \theta_{\rm T}^{\phantom{T}}$, where $D_{\rm LT}^{\phantom{T}} = - \left[ J_{\rm LL}^{\phantom{T}} \right]^{-1} J_{\rm LT}^{\phantom{T}}$.

\par Thus, the effective interactions between the ${\rm T}$ spins will have the following form:
\noindent
\begin{equation}
\tilde{J}_{\rm TT} = J_{\rm TT} -  J_{\rm TL} \left[ J_{\rm LL} \right]^{-1} J_{\rm LT}
\label{eq:jTT}
\end{equation}
\noindent and
\noindent
\begin{equation}
\tilde{d}^{z}_{\rm TT} = d^{z}_{\rm TT} +  D_{\rm TL}^{\phantom{T}}d^{z}_{\rm LT} + d^{z}_{\rm TL} D_{\rm LT}^{\phantom{T}} + D_{\rm TL}^{\phantom{T}} d^{z}_{\rm LL} D_{\rm LT}^{\phantom{T}}.
\label{eq:dzTT}
\end{equation}
\noindent In comparison with the bare interactions, $J_{\rm TT}$ and $d^{z}_{\rm TT}$, the parameters $\tilde{J}_{\rm TT}$ and $\tilde{d}^{z}_{\rm TT}$ acquire the additional contributions related to the magnetic polarization of the ligand sites.

\par Somewhat similar technique was employed by Mryasov~\textit{at al.} in order to explain the unusual temperature dependence of the magnetic anisotropy energy in the ordered FePt alloy~\cite{Mryasov}.

\section{\label{sec:Results} Applications}
\par Below we apply these techniques for the analysis of DM interactions in CrI$_3$ and CrCl$_3$. Particularly, CrI$_3$ is regarded as the prominent two-dimensional (2D) van der Waals ferromagnet~\cite{CrI3_Huang_Nature} as well as the suitable testbed material for studying the effects of the ligand states on interatomic exchange interactions~\cite{PRB2019}. The bulk CrI$_3$ and CrCl$_3$ crystallize in the centrosymmetric $R\overline{3}$ structure. More details can be found in Refs.~\cite{CrCl3str,CrI3str}.

\par We consider several scenarios of the inversion symmetry breaking, which lead to the emergence of DM interactions (see Fig.~\ref{fig.layers}):
\noindent
\begin{figure*}[t]
\begin{center}
\includegraphics[width=15.0cm]{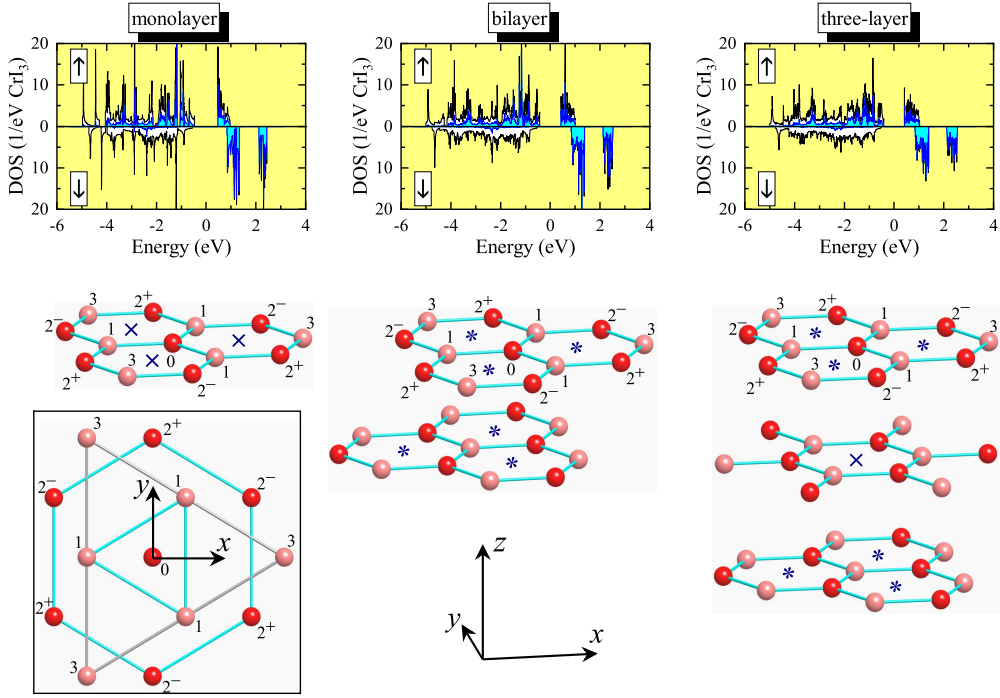}
\end{center}
\caption{Monolayer, bilayer, and three-layer CrI$_3$ with the notations of first, second, and third coordination spheres of the Cr atoms relative to the central site $0$. The inversion centers, transforming the layered structures to themselves, are denoted by $\times$. In addition to them, the spacial inversion about some centers, denoted by $*$, can be combined with the shift connecting the top and bottom layer. Two Cr sublattice (I and II), which are formed in the bulk, are denoted by different colors. Top panel shows the total and partial Cr $3d$ densities of states (white and cyan areas, respectively). The zero energy is in the middle of the gap between occupied and unoccupied states.
}
\label{fig.layers}
\end{figure*}
\noindent
\begin{itemize}
\item[(i)] Some DM interactions can emerge already in the pristine centrosymmetric $R\overline{3}$ structure~\cite{Kvashnin2020}. This structure has two Cr sublattices, which can be transformed to each other by the spacial inversion about the centers of the hexagons. Obviously, all DM interactions between such sublattices will vanish. However, the Cr atoms in each of the sublattices (which are shown by the same colors in Fig.~\ref{fig.layers}) are not connected by the inversion. Therefore, the DM interaction within each sublattice are allowed by the $R\overline{3}$ symmetry. For instance, such interactions will take place between 2nd neighbors in the honeycomb plane;
\item[(ii)] The inversion symmetry can be broken by the electric field. For these purposes we consider the perpendicular electric field $\boldsymbol{E}=(0,0,E)$ applied to monolayer (1L) CrI$_3$, which is simulated by shifting the on-site energies of the tight-binding Hamiltonian by $-\boldsymbol{E} \cdot \boldsymbol{R}_{i}$~\cite{PCCP2019}. Without the electric field, 1L-CrI$_3$ is centrosymmetric and the DM vectors have the same symmetry properties as in the bulk.
\item[(iii)] The inversion symmetry is broken at the surface. We consider the effects of such inversion symmetry breaking in the bilayer (2L) and three-layer (3L) CrI$_3$. The 3L-CrI$_3$ is centrosymmetric, where the inversion centers are located in the middle layer. Obviously, the inversion about similar centers in the surface layers is no longer possible. Nevertheless, such inversion operation in the two-dimensional (2D) systems can still be combined with the shift connecting the top and bottom layers, thus, imposing additional symmetry constraints on the form of the DM interactions.
\end{itemize}

\par For practical purposes, we use the linear muffin-tin orbital method (LMTO) in the atomic-spheres approximation (ASA)~\cite{LMTO1,LMTO2} for the electronic structure calculations in the local spin-density approximation (LSDA)~\cite{VWN} with the experimental parameters of the crystal structure for the bulk CrCl$_3$~\cite{CrCl3str} and CrI$_3$~\cite{CrI3str}. The tight-binding Hamiltonian for the bulk systems was constructed using the projector-operator technique~\cite{WannierRevModPhys,JPCMreview}. Then, the same parameters were used to simulate the 2D $n$L-CrI$_3$ systems. Such approach has obvious limitations and is not aiming at the overwhelming quantitative description of the surface states, which should include the structural relaxation and, possibly, the correlations effects beyond LSDA~\cite{Kvashnin2022}. Nevertheless, we expect it to gives a clear idea about the microscopic origin of DM interactions at the surface of CrI$_3$. The response tensor was calculated on the mesh of $10 \times 10 \times 10$ ($10 \times 10 \times 1$) $\boldsymbol{q}$-points and the $\boldsymbol{k}$-space integration was performed on the mesh of $20 \times 20 \times 20$ ($54 \times 54 \times 1$) points in the first Brillouin zone for the bulk (2D) materials.

\par The DM vectors in the bonds, which are transformed to each other by the threefold rotations, can be presented in the form $\boldsymbol{d} = \big( d^{\perp}\cos [\phi+\psi], d^{\perp}\sin[\phi+\psi], d^{z} \big)$, where $\phi$ is the azimuthal angle specifying the direction of the bond in the $xy$ plane and $\psi$ specifies the direction of the DM vector in the plane relative to this bond. Thus, the DM vector is perpendicular to the bond if $\psi = \pm 90^{\circ}$ and parallel to it if $\psi = 0$ or $180^{\circ}$.

\subsection{\label{sec:symmetry} Basic symmetry properties of Dzyaloshinskii-Moriya interactions}
\par When the CrI$_3$ layer contains the inversion center transforming two Cr sublattices (${\rm I}$ and ${\rm II}$) to each other, the DM interactions in these sublattices are related by $\boldsymbol{d}_{\rm II}({\bf R}) = -\boldsymbol{d}_{\rm I}({\bf R})$. This property holds for each pair of sites separated by the vector ${\bf R}$~\cite{PRB2013}. Such a situation is realized, for instance, for the 2nd neighbor interactions in the honeycomb plane. Furthermore, the same symmetry property for the DM interactions between the sublattices (for instance, in the 1st and 3rd coordination spheres within the layer) yields $\boldsymbol{d}({\bf R}) = -\boldsymbol{d}({\bf R})$, meaning that they all vanish. Such a situation is realized in the bulk, 1L-CrI$_3$ and the middle layer of 3L-CrI$_3$. If the inversion symmetry is broken (either by the electric field or at the surface), the DM interactions between the sublattices will emerge. Moreover, the sublattices become inequivalent, so that the DM interactions will acquire the following form: $\boldsymbol{d}_{\rm I,II}({\bf R}) = \bar{\boldsymbol{d}}({\bf R}) \pm \Delta \boldsymbol{d}({\bf R})$, where $\bar{\boldsymbol{d}}({\bf R}) = \frac{1}{2}\{ \boldsymbol{d}_{\rm I}({\bf R}) - \boldsymbol{d}_{\rm II}({\bf R}) \}$ and $\Delta \boldsymbol{d}({\bf R}) = \frac{1}{2}\{ \boldsymbol{d}_{\rm I}({\bf R}) + \boldsymbol{d}_{\rm II}({\bf R}) \}$. The isotropic exchange interactions are also expected to change as $J_{\rm I,II}({\bf R}) = \bar{J}({\bf R}) \pm \Delta J({\bf R})$, where $\bar{J}({\bf R}) = \frac{1}{2}\{ J_{\rm I}({\bf R}) + J_{\rm II}({\bf R}) \}$ and $\Delta \bar{J}({\bf R}) = \frac{1}{2}\{ J_{\rm I}({\bf R}) - J_{\rm II}({\bf R}) \}$.

\subsection{\label{sec:DMgeneral} General tendencies of Dzyaloshinskii-Moriya interactions}
\par First, we review general tendencies of the exchange interactions, obtained using the exact Eqs.~(\ref{eq:jexactM}) and (\ref{eq:dexactM}) and taking into account the contributions of the ligand states as described in Sec.~\ref{sec:downfolding}. The results are summarized in Table~\ref{tab:Msummary} for all considered systems.
\noindent
\begin{table*}[t]
\caption{Isotropic and DM interactions (in meV) in the 1st, 2nd, and 3rd coordination spheres of the CrI$_3$ and CrCl$_3$ planes in the bulk, monolayer (1L), bilayer (2L), and three-layer (3L) CrI$_3$ (where the top and middle planes are denoted, respectively, by t and m). The notation $+E$ stands for calculations in the external electric field $E=0.1$ V/\AA. $\psi$ is the phase (in degrees) specifying the direction of the DM vector relative to the bond. The parameters were obtained by using the exact approach and take into account the contributions of the ligand states. The parameters for the 2nd coordination sphere were averaged over two sublattices (see text for details). The positive (negative) $d^{z}$ is realized in the bonds $0$-$2^{+}$ ($0$-$2^{-}$) (see Fig.~\ref{fig.layers} for the notations of the atomic positions).}
\label{tab:Msummary}
\begin{ruledtabular}
\begin{tabular}{lcccccccccccc}
                        & \multicolumn{4}{c}{1st coordination}             & \multicolumn{4}{c}{2nd coordination}                                & \multicolumn{4}{c}{3rd coordination} \\
\hline
     system             & $J$    & $d^{\perp}$ & $\psi$ & $d^{z}$ & $\bar{J}$ & $\bar{d}^{\perp}$ & $\bar{\psi}$ & $\bar{d}^{z}$ & $J$     & $d^{\perp}$ & $\psi$          & $d^{z}$           \\
\hline
bulk-CrCl$_3$           & $2.27$ & $-$         & $-$    & $-$     & $0.32$    & $0.01$            & $24$         & $\pm 0.02$    & $-0.35$ & $-$         & $-$             & $-$               \\
bulk-CrI$_3$            & $1.53$ & $-$         & $-$    & $-$     & $0.77$    & $0.09$            & $0$          & $\pm 0.28$    & $-0.50$ & $-$         & $-$             & $-$               \\
1L-CrI$_3$              & $0.34$ & $-$         & $-$    & $-$     & $1.14$    & $0.12$            & $0$          & $\pm 0.30$    & $-0.65$ & $-$         & $-$             & $-$               \\
1L-CrI$_3$ $+E$         & $0.20$ & $0.82$      & $89$   & $0.13$  & $1.19$    & $0.12$            & $0$          & $\pm 0.30$    & $-0.63$ & $0.01$      & $72$            & $-0.11$           \\
2L-CrI$_3$              & $0.99$ & $0.36$      & $91$   & $0.11$  & $0.96$    & $0.10$            & $-22$        & $\pm 0.29$    & $-0.57$ & $0.01$      & $135$           & $\phantom{-}0.02$ \\
3L-CrI$_3$ (t)          & $0.99$ & $0.37$      & $91$   & $0.11$  & $0.95$    & $0.09$            & $0$          & $\pm 0.29$    & $-0.57$ & $0.01$      & $131$           & $\phantom{-}0.02$ \\
3L-CrI$_3$ (m)          & $1.52$ & $-$         & $-$    & $-$     & $0.78$    & $0.09$            & $0$          & $\pm 0.28$    & $-0.50$ & $-$         & $-$             & $-$
\end{tabular}
\end{ruledtabular}
\end{table*}

\par As expected from the general symmetry considerations (Sec.~\ref{sec:symmetry}), the DM interactions in the 1st and 3rd coordination sphere of the bulk, 1L-CrI$_3$, and the middle layer of 3L-CrI$_3$ identically vanish. These interactions can be induced only by breaking the inversion symmetry either by the electric field or at the surface layers of 2L-CrI$_3$ and 3L-CrI$_3$. In CrI$_3$, the DM interactions can be strong and comparable with isotropic exchange interactions. Such situation is realized, for instance, in the 2nd coordination sphere of all considered CrI$_3$ systems and also in the 1st coordination sphere, if these DM interactions are allowed by the symmetry. The DM interactions in the 3rd coordination sphere are considerably weaker, with some exception of 1L-CrI$_3$ in the electric field yielding sizable $d^{z}$. In order to appreciate the role of the ligand states, it is instructive to compare the results for the bulk CrI$_3$ and CrCl$_3$. The Cl atoms are lighter and, therefore, have weaker SO coupling. The DM interactions in CrCl$_3$ are reduced drastically in comparison with those in CrI$_3$, suggesting that the ligand states play a crucial role in microscopic processes responsible for these interactions, similar to the magnetic anisotropy~\cite{Lado2017}. This point will be further elaborated below.

\par
\begin{figure}[b]
\begin{center}
\includegraphics[width=7.0cm]{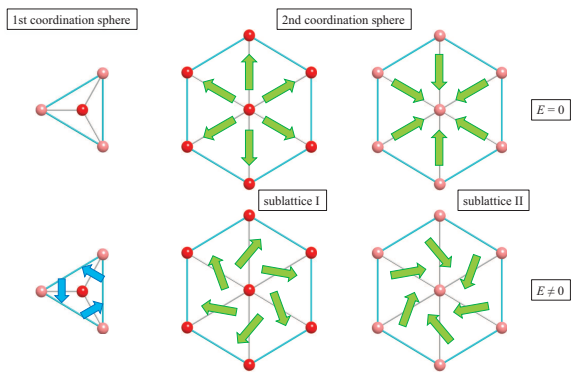}
\end{center}
\caption{Cartoon picture explaining distribution of the Dzyaloshinskii-Moriya vectors in the 1st coordination sphere and two Cr sublattices in the 2nd coordination sphere of 1L-CrI$_3$ with (bottom) and without (top) the electric field applied perpendicular to the layer.
}
\label{fig.1L-CrCl3}
\end{figure}

 For the 2nd coordination sphere, Table~\ref{tab:Msummary} shows the parameters averaged over two Cr sublattices. If the layer contains the inversion centers, these are the true interactions operating in the layer. If not, two sublattices become inequivalent and will generally be characterized by different interactions. Nevertheless, the deviations from the averaged values seem to be small. For instance, in the 1L-CrI$_3$ case with the electric field, we obtain $\Delta J = -0.01$ meV, and $\Delta d^{\perp} \approx \Delta d^{z} \approx 0$. The most interesting consequence of the electric field is the behavior of phases $\psi_{\rm I,II}$. The averaged phase $\bar{\psi}$ vanishes, meaning that the averaged DM vector is parallel to the bond, as without field. However, this is the result of cancellation, whereas the individual phases in the sublattices I and II are finite and can be evaluated as $\psi_{\rm I,II} = \pm 40^{\circ}$. Thus, the application of the electric field leads to the additional rotations of the DM vectors in the $xy$ plane away from the bonds, which for the sublattices ${\rm I}$ and ${\rm II}$ occur in the antiphase, as shown in Fig.~\ref{fig.1L-CrCl3}.
\noindent

\par Kvashnin \textit{et al.} reported results of MFT calculations for the bare exchange parameters in the series of Cr$X_3$ compounds ($X=$ Cl, Br, I)~\cite{Kvashnin2020}. Particularly, they have confirmed that the main DM interaction, which is allowed by the symmetry in the honeycomb plane, is the one in the 2nd coordination sphere. According to their calculations, the length of the corresponding DM vector, $|\boldsymbol{d}| = \sqrt{(d^{\perp})^{2} + (d^{z})^{2}}$, changes from about $0.15$ to $0.20$ meV when going from the bulk CrI$_3$ to 1L-CrI$_3$. These values should be compared with $|\boldsymbol{d}| =$ $0.28$ and $0.31$ meV obtained in the present work within bare MFT (see Supplemental Material~\cite{SM} for details). Taking into account the sensitivity of the DM parameters to the structural relaxation, which was considered in Ref.~\cite{Kvashnin2020} but not in the present work, the agreement is reasonable. Particularly, the structural relaxation should explain the larger difference of the parameters $|\boldsymbol{d}|$ between bulk CrI$_3$ to 1L-CrI$_3$. Furthermore, Kvashnin \textit{et al.} used the generalized gradient approximation~\cite{PBE} implemented in the full-potential LMTO code~\cite{FPLMTO} (and VASP code~\cite{VASP} for the structural optimization), while we use LSDA and the ASA-LMTO code, which result in the additional difference of the exchange parameters, as was elaborated in details in Ref.~\cite{PRB2019}. There is also an overall agreement between values of isotropic exchange interactions: for instance, for the 1st neighbor interactions in CrI$_3$ and 1L-CrI$_3$ Kvashnin \textit{et al.} reported $J \approx$ $2.2$ and $1.2$ meV, respectively, which should be compared with $J \approx$ $3.0$ and $1.8$ meV, obtained in the present work~\cite{SM}.

\par Next, we compare abilities of different techniques for calculating isotropic and DM interactions. For these purposes we consider the representative example of 1L-CrI$_3$ in the electric field, which nicely captures the main tendencies. The results are summarized in Table~\ref{tab:comparison}.
\noindent
\begin{table*}[t]
\caption{Isotropic and DM interactions (in meV) in the 1st, 2nd, and 3rd coordination spheres of monolayer CrI$_3$ in the external electric field $E=0.1$ V/\AA~as obtained in MFT and the exact approach. $3d$ denotes the bare interactions involving only the Cr $3d$ states. $3d$$+$${\rm L}$ denotes the same interactions, which were additionally corrected by interactions with the ligand states by using the downfolding method. $\psi$ is the phase (in degrees) specifying the direction of the DM vector relative to the bond. The parameters for the 2nd coordination sphere were averaged over two sublattices (see text for details). The positive (negative) $d^{z}$ is realized in the bonds $0$-$2^{+}$ ($0$-$2^{-}$) (see Fig.~\ref{fig.layers} for the notation of the atomic sites).}
\label{tab:comparison}
\begin{ruledtabular}
\begin{tabular}{lrccccccccccc}
                        & \multicolumn{4}{c}{1st coordination}                         & \multicolumn{4}{c}{2nd coordination}                                 & \multicolumn{4}{c}{3rd coordination}               \\
\hline
     method             & $J$~~~   & $d^{\perp}$ & $\psi$ & $d^{z}$           & $\bar{J}$         & $\bar{d}^{\perp}$ & $\bar{\psi}$ & $\bar{d}^{z}$ & $J$     & $d^{\perp}$ & $\psi$ & $d^{z}$           \\
\hline
MFT, $3d$               & $1.69$   & $0.81$      & $89$   & $\phantom{-}0.11$ & $\phantom{-}1.21$ & $0.12$            & $0$          & $\pm 0.29$    & $-0.56$ & $0$         & $70$   & $-0.10$           \\
MFT, $3d$$+$${\rm L}$   & $-1.13$  & $0.78$      & $89$   & $\phantom{-}0.13$ & $\phantom{-}1.37$ & $0.12$            & $0$          & $\pm 0.27$    & $-0.53$ & $0$         & $39$   & $-0.10$           \\
exact, $3d$             & $-29.69$ & $0.17$      & $90$   & $-0.07$           & $-1.54$           & $0.08$            & $0$          & $\pm 0.11$    & $-1.79$ & $0.06$      & $88$   & $-0.16$           \\
exact, $3d$$+$${\rm L}$ & $0.20$   & $0.82$      & $89$   & $\phantom{-}0.13$ & $\phantom{-}1.19$ & $0.12$            & $0$          & $\pm 0.30$    & $-0.63$ & $0.01$      & $72$   & $-0.11$
\end{tabular}
\end{ruledtabular}
\end{table*}
\noindent First, we note that MFT and exact technique provide very different description for isotropic exchange interactions. Formally, MFT is able to reproduce the ferromagnetic character of isotropic interactions in the 1st and 2nd coordination spheres, but only if we consider bare interactions and ignore the ligand contributions. The ligand states worsen the situation by making the 1st neighbor coupling antiferromagnetic. The description based on the exact approach appears to be more consistent: the Cr $3d$ states alone lead to the antiferromagnetic coupling, while the ferromagnetism in the 1st and 2nd coordination spheres is solely related to the ligand states. Similar tendency was obtained in the bulk CrI$_3$~\cite{PRB2021}, in agreement with the Goodenough-Kanamori-Anderson rules for the $90^{\circ}$ exchange~\cite{Kanamori_GKA}: the ferromagnetism in this cases arises from the Hund coupling on the ligand sites, which is incorporated in the second term of Eq.~(\ref{eq:jTT})~\cite{Arxiv2022}.

\par The behavior of the DM interactions is different. In fact, there are several ways how the SO coupling on the ligand states contributes to the values of DM interactions between the transition-metal sites. For the bare interactions, the ligand states play a role of an effective medium, which connects the $3d$ orbitals at different transition-metal sites by means of the hybridization effects. Naively, since the ligand $p$ states are almost fully occupied, one could expect that the SO coupling associated with this group of states should not be primarily important for the bare DM interactions between the transition-metal sites, which are given by the first term of Eq.~(\ref{eq:dzTT}). Nevertheless, our results show that the SO coupling on the heavy ligand sites has a profound effect on these bare DM interactions. This can be seen by artificially switching off the SO coupling on the ligand sites, which leads to the sharp drop (by an order of magnitude) of the DM interactions, as illustrated in Supplemental Material~\cite{SM}. This is also consistent with the results of Kvashnin \textit{et al.}~\cite{Kvashnin2020}, which show that the DM parameters sharply decrease in the direction CrI$_3$ $\rightarrow$ CrBr$_3$ $\rightarrow$ CrCl$_3$, indicating that the magnitude of the bare DM interactions between the Cr sites is controlled by the SO coupling on the ligand sites.

\par Then, similar to the isotropic exchange, the additional contribution to the DM interaction arises from the magnetic polarization of the ligand states, which is described by three remaining terms in Eq.~(\ref{eq:dzTT}). These contributions appear to be relatively weak in MFT and only slightly correct the values of bare interactions. However, in the exact approach, this correction is much more important: for the nearest neighbors, it strengthens $d^{\perp}$ from $0.17$ to $0.82$ meV and changes the sign of $d^{z}$. Appreciable changes are seen also in the 2nd and 3rd coordination spheres. As expected, the correction is small when the SO interactions is considered only on the Cr sites~\cite{SM}.

\par Somewhat phenomenologically, we note that the MFT approach for the DM interactions works surprisingly well: when we consider the exact technique and take into account the contributions of the ligand sites (the $3d$$+$${\rm L}$ scheme), the obtained DM parameters are well consistent with the MFT results. This seems to be a general trend, at least for this particular type of compounds, which applies not only to 1L-CrI$_3$ (Table~\ref{tab:comparison}), but also to other considered systems~\cite{SM}. Nevertheless, we would like to caution again that the same rule does not apply to isotropic interactions.

\par One intuitive reason why MFT does a better job for the DM interactions is that in this case it operates with small matrix elements of the one-electron Hamiltonian, which emerge in the first order of the SO coupling and, therefore, enforces the strong coupling limit (the large interatomic xc splitting in comparison with other physically relevant quantities), where MFT becomes exact~\cite{PRB2021}.

\par Finally, we compare our results with experimental data. First, we have realized that our previous calculations for the bulk CrCl$_3$ and CrI$_3$~\cite{PRB2021} contained a mistake because we treated all $J_{\mu \nu} (\boldsymbol{q})$ as real variables. However, this is true only for intrasublattice interactions (for instance, occurring in the 2nd coordination sphere in the layer), where the intersublattice interactions (in the 1st and 3rd coordination spheres) across the inversion center can be complex. The corrected parameters in the plane are reported in Table~\ref{tab:Msummary}. All set of isotropic exchange interactions, obtained in the exact approach and taking into account the contributions of the ligand states for CrI$_3$ (CrCl$_3$) is is $J_{1}$$=$ $1.53$ ($2.27$) meV, $J_{2}$$=$ $0.76$ ($0.28$) meV, $J_{3}$$=$ $0.77$ ($0.32$) meV, $J_{4}$$=$ $0.84$ ($0.12$) meV, $J_{5}$$=$ $0.55$ ($0.13$) meV, and $J_{6}$$=$ $-0.50$ ($-0.35$) meV in the notations of Ref.~\cite{PRB2021}, where $J_{1}$, $J_{3}$, and $J_{6}$ are the inplane interactions (corresponding to the 1st, 2nd, and 3rd coordination spheres), whereas $J_{2}$, $J_{4}$, and $J_{5}$ are the exchange interactions between the planes. Moreover, $J_{4}$ and $J_{5}$ contribute to the spin-wave dispersion in the plane. The corresponding Curie temperature, evaluated within random phase approximation~\cite{PRM2019}, is $66$ ($33$) K, which agree with the experimental value of $61$ ($17$) K for CrI$_3$ (CrCl$_3$)~\cite{ChenPRX,McGuirePRM}. The use of MFT for the exchange interactions worsens the agreement~\cite{SM}.

\par The theoretical spin-wave dispersion, $\omega(\boldsymbol{q})$, is shown in Fig.~\ref{fig.SW} and the details of calculations are explained in the Supplemental Material~\cite{SM}.
\noindent
\begin{figure}[t]
\begin{center}
\includegraphics[width=8.6cm]{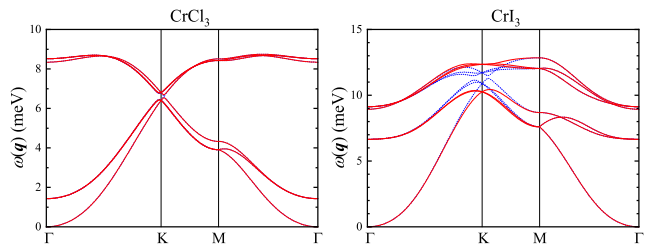}
\end{center}
\caption{Spin-wave dispersion as obtained for the bulk CrCl$3$ (left) and CrCl$3$ (left) using the exact approach for the exchange parameters, which also takes into account the contributions of the ligand states, with (solid) and without (dotted) the Dzyaloshinskii-Moriya interactions. The calculations are performed for the hexagonal cell, where six branches of $\omega(\boldsymbol{q})$ correspond to six magnetic Cr sublattices.}
\label{fig.SW}
\end{figure}
\noindent The quantitative agreement with the experimental data for CrI$_3$~\cite{ChenPRX} is far from being perfect. Nevertheless, we would like to comment on the origin of the spin-wave gap at the Dirac K point, which is actively discussed in the literature~\cite{ChenPRX,LeePRL,Costa,Olsen,KeKatsnelson}. First, we note that the gap can be open already by isotropic interplane interactions $J_{4}$ and $J_{5}$, in agreement with Ref.~\cite{KeKatsnelson}. However, the effect is not particularly strong (but stronger than in CrCl$_3$, where $J_{4}$ and $J_{5}$ are weaker). The strong DM interactions, $d^{z}_{ij}$, mainly those operating in the 2nd coordination sphere of the CrI$_3$ plane, further widen the gap at the K point, as was originally proposed in Ref.~\cite{ChenPRX}. For comparison, the effect of DM interactions in CrCl$_3$ is nearly negligible. The theoretical gap in CrI$_3$, $\sim 2$ meV, is about two times smaller than the experimental one~\cite{ChenPRX,LeePRL}. Nevertheless, we consider such agreement as encouraging. First, the effect of the SO coupling (and, therefore, DM interactions) can be further enhanced by the on-site Coulomb interactions~\cite{PRB2014}, which were not considered in the present work. Then, the contributions of the ligand states rely on the values of Stoner parameters, which arrear to be not well defined~\cite{Arxiv2022}. In the present work, we use for these purposes the sum rule, but it is possible than other definitions can improve the agreement with the experimental data~\cite{Arxiv2022}.

\section{\label{sec:Alternative} Alternative methods}

\subsection{\label{sec:mixedPT} Cycloidal spirals in the long wavelength limit and mixed perturbation theory}
\par
\begin{figure}[t]
\begin{center}
\includegraphics[width=8.6cm]{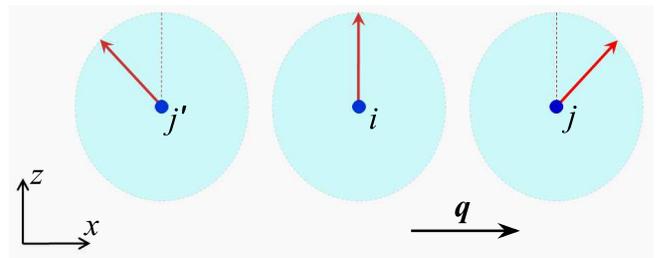}
\end{center}
\caption{Cycloidal spin spiral, propagating along $x$ and rotating in the $zx$ plane, which can be alternatively used in order to calculate $y$ component of the DM vector $\boldsymbol{d}$ perpendicular to the plane.}
\label{fig.cartoon2}
\end{figure}

 The choice of the conical spin configurations (Fig.~\ref{fig.cartoon3}) for calculating the DM interactions is not unique. Alternatively, one can consider the cycloidal spin spiral with $\boldsymbol{e}_{i}=(\sin \boldsymbol{q} \cdot \boldsymbol{R}_{i},0,\cos \boldsymbol{q} \cdot \boldsymbol{R}_{i})$ (see Fig.~\ref{fig.cartoon2}) and treat small rotations of the xc field in the limit $\boldsymbol{q} \to 0$ within first order perturbation theory~\cite{Freimuth}.
\noindent
\noindent In this sense, this method is similar to the MFT method by Liechtenstein \textit{et al.}~\cite{LKAG1987} as both of them deal with the rotations of the xc field and ignore the constraining field, which is required in order to fix the directions of spins~\cite{BrunoPRL2003,Streib,PRB2021}.

\par Then, if we treat the SO coupling as another perturbation, the corresponding energy change can be presented in the form of pairwise interactions, where the rotation of the xc field at the site $i$ is combined with the SO interaction at the site $j$ (and vice versa)~\cite{PRL96}. In the noncentrosymmetric bond $ij$, the SO interaction on the sites $j$ will create a force rotating the spin on the site $i$, which can be related to the DM interaction~\cite{PRL96,PRB2014}. The method was applied to LaMnO$_3$~\cite{PRL96}, by considering the rotations of the xc field and the SO interactions only on the magnetic Mn sites. Then, it was further generalized to include the contributions related to the polarization of the electron system, also emerging in the first order of the SO coupling~\cite{PRB2014}. All these calculations take into account the SO coupling only on the magnetic sites and ignore the contributions of the ligand sites. The results for CrI$_3$ are summarized in the Supplemental Material~\cite{SM}, where we also compare the mixed perturbation theory~\cite{PRL96} with the MFT based approach considered in Sec.~\ref{sec:MFT}. As long as we consider the SO interactions only on the magnetic Cr sites, these two methods provide a consistent description. However, it is obvious that such approach will severely underestimate of the DM interaction parameters in the case of CrI$_3$.

\subsection{\label{sec:SpinCurrent} Spin-current model}
\par
\begin{figure}[t]
\begin{center}
\includegraphics[width=8.6cm]{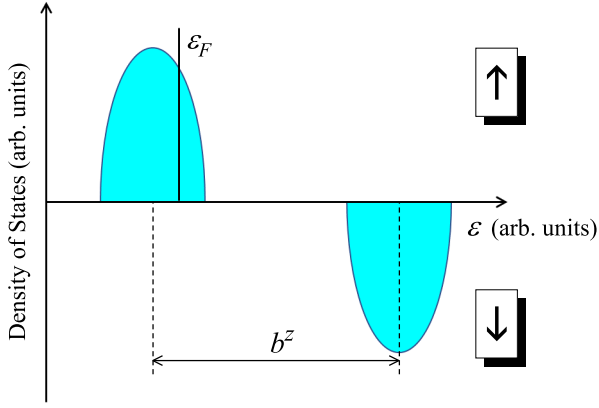}
\end{center}
\caption{Schematic density of states: partially occupied $\uparrow$-spin band and unoccupied $\downarrow$-spin band split by the exchange-correlation field $b^{z}$. The double exchange model corresponds to the limit $b^{z} \to \infty$.}
\label{fig.cartoon4}
\end{figure}

 In this sections we consider the spin-current model for the DM interactions and show how it can be derived from the more general Eq.~(\ref{eq:dijzMFT}) in the strong coupling limit $b^{z}_{i} \to \infty$, which justifies the use of MFT~\cite{PRB2021}. Furthermore, the limit $b^{z} \to \infty$ is equivalent to the assumption of strong intraatomic exchange coupling made in the spin-current model~\cite{Kikuchi}. For the practical purposes, it is convenient to adopt the one-electron Hamiltonian in the form
\noindent
\begin{displaymath}
\hat{H}^{\sigma}_{ij} = \left\{
\begin{array}{ll}
\hat{t}^{\sigma}_{ij}                           & \textrm{if $i \ne j$} \\
\hat{b}^{z}_{\phantom{i}}\delta_{s, \downarrow} & \textrm{if $i = j$}
\end{array}
\right. ,
\end{displaymath}
\noindent where all states are additionally shifted upwards by $\frac{1}{2} \hat{b}^{z}$ and $\hat{b}^{z}$ is assumed to be the same for all magnetic sites (see Fig.~\ref{fig.cartoon4}). Moreover, all affects associated with the SO interaction are incorporated into the transfer integrals $\hat{t}^{\sigma}_{ij}$, as it is typically assumed in the model analysis~\cite{Katsnelson_DM}.
\noindent

\par If the $\downarrow$-spin states are unoccupied (see Fig.~\ref{fig.cartoon4}), the leading term of $\hat{G}^{\downarrow}_{ij}$ in the occupied part of the spectrum for $i \ne j$ is given by $\hat{G}^{\downarrow}_{ij} \approx (\hat{b}^{z})^{-1} \hat{t}^{\downarrow}_{ij} (\hat{b}^{z})^{-1}$. By substituting it to Eq.~(\ref{eq:dijzMFT}), one obtains the following expression for the DM interactions:
\noindent
\begin{equation}
d_{ij}^{z} = \frac{1}{4 \pi} {\rm Im} \, i \int_{- \infty}^{\varepsilon_F} d \varepsilon \, {\rm Tr}_{L} \left\{ \hat{G}^{\uparrow}_{ij}(\varepsilon) \hat{t}^{\downarrow}_{ji} - \hat{t}^{\downarrow}_{ij} \hat{G}^{\uparrow}_{ji}(\varepsilon) \right\}.
\label{eq:dijzDE1}
\end{equation}

\par To the first order in the SO coupling, $\hat{t}^{\downarrow}$ and $\hat{G}^{\uparrow}$ in Eq.~(\ref{eq:dijzDE1}) can be rearranged as: $\hat{t}^{\downarrow} = \hat{t}^{0} -i \hat{t}^{z}$ and $\hat{G}^{\uparrow} = \hat{G}^{0} + i\hat{G}^{z}$. The transfer integrals can be generally written as $\hat{\tt{t}} = t^{0} \hat{\mathbb{1}} + i \boldsymbol{t}^{\phantom{0}} \cdot \hat{\boldsymbol{\sigma}}$~\cite{1orbital}, where $\hat{\mathbb{1}}$ is the unity matrix, $\hat{\boldsymbol{\sigma}}$ is the vector of Pauli matrices, and $\boldsymbol{t}^{\phantom{0}} = (t^{x},t^{y},t^{z})$ is induced by the SO interaction in the noncentrosymmetric bonds (otherwise, $\hat{\tt{t}} = t^{0} \hat{\mathbb{1}}$). $\hat{G}^{z}$ can be regarded as the perturbation to the Green function induced by $\hat{t}^{z}$. Then, we arrive at the following expression for $d_{ij}^{z}$:
\noindent
\begin{equation}
d_{ij}^{z} = \frac{1}{2 \pi} {\rm Im} \, \int_{- \infty}^{\varepsilon_F} d \varepsilon \, {\rm Tr}_{L} \left\{ \hat{G}^{0}_{ij}(\varepsilon) \hat{t}^{z}_{ji} - \hat{G}^{z}_{ij}(\varepsilon) \hat{t}^{0}_{ji} \right\}.
\label{eq:dijzDE2}
\end{equation}

\par In order to relate this expression to the spin current, one should recall that the latter is defined as the anticommutator of $\hat{\tt{t}}$ and $\sigma^{\alpha}$: $j_{s}^{\alpha} \sim \left( \hat{\tt{t}} \sigma^{\alpha} + \sigma^{\alpha} \hat{\tt{t}} \right)$~\cite{Katsnelson_DM,Kikuchi}. Then, one can say that $d_{ij}^{z}$ in Eq.~(\ref{eq:dijzDE2}) is induced by the spin current $j_{s}^{z}$ and the change of the Green function caused by the same spin current, in agreement with Refs.~\cite{Katsnelson_DM,Kikuchi}. Furthermore, $j_{s}^{z}$ is proportional to only $t^{z}$ and does not depend on $t^{x}$ and $t^{y}$, which follows directly from the anticommutation properties of the Pauli matrices. Therefore, in order to calculate $j_{s}^{z}$ ($d^{z}_{ij}$), it is sufficient to know only the change of the electronic structure induced by the SO interaction separately for $\sigma$$=$ $\uparrow$ and $\downarrow$, which in the spin-current model is described by $t^{z}$. The coupling between $\sigma$$=$ $\uparrow$ and $\downarrow$, which is described by $t^{x}$ and $t^{y}$, can be neglected. This is totally consistent with the assumption made by Sandratskii in his seemingly different approach~\cite{Sandratskii}.

\par Nevertheless, it should be understood that the spin-current expression (\ref{eq:dijzDE2}) is just the limiting case of the more general expression given by Eq.~(\ref{eq:dijzMFT}). For the isotropic interactions (\ref{eq:JijMFT}), the same limit $b^{z} \to \infty$ yields the double exchange parameters~\cite{PRL1999}:
\noindent
\begin{equation}
J_{ij} = \frac{1}{2 \pi} {\rm Im} \, \int_{- \infty}^{\varepsilon_F} d \varepsilon \, {\rm Tr}_{L} \left\{ \hat{G}^{0}_{ij}(\varepsilon) \hat{t}^{0}_{ji} \right\}.
\label{eq:JDE}
\end{equation}
\noindent Typically, the double exchange limit ($b^{z} \to \infty$) is insufficient and the correct description of the exchange interactions in realistic materials requires other contributions emerging in the higher orders of $\left( b^{z} \right)^{-1}$~\cite{PRB2015a}. Particularly, the double exchange alone fails to account for antiferromagnetic interactions in insulating materials and vanishes if the electronic states are half-filled. The key microscopic mechanism resulting in the antiferromagnetic coupling is know to be superexchange~\cite{Anderson}. The corresponding parameters can be also derived from Eq.~(\ref{eq:JijMFT}), but in the first order of $\hat{t}^{0}_{ji}(b^{z})^{-1}$. The expression (\ref{eq:JDE}) does not depend on $b^{z}$ and does not take into account the effects associated with the superexhange mechanism. Similar limitations are expected for the DM interactions in the spin-current model.

\section{\label{sec:summary} Summary}
\par We discussed various techniques for calculating antisymmetric DM interactions on the basis of first-principles electronic structure calculations and showed how these, seemingly different approaches, can be unified. First, we corroborated the idea of Sandratskii~\cite{Sandratskii} stating that, in order to calculate $z$ component of the DM vector, it is sufficient to know only the change of the electronic structure caused by the spin-diagonal part of the SO interaction, while all complications related to the off-diagonal elements can be neglected. This assumption, which becomes exact to the leading order in the SO coupling, greatly simplifies the calculations and allowed us to derive a transparent expression for $d^{z}_{ij}$ in the framework of MFT for the infinitesimal rotations of the xc fields~\cite{LKAG1987}. Basically, this $d^{z}_{ij}$ can be computed at the same cost as the isotropic exchange interaction $J_{ij}$. Then, we switched to the more rigorous technique, which is formulated in terms of the inverse response function and cures the main limitations of MFT.  Furthermore, we showed how on can downfold the ligand spins by incorporating their effect into the effective exchange interaction between the localized spins.

\par The abilities of these techniques were demonstrated for CrCl$_3$ and CrI$_3$. Particularly, we have shown that, even in the centrosymmetric bulk structure, the DM interactions emerge between Cr atoms located in the same sublattice. Then, we considered how the inversion symmetry breaks in the surface layer of CrI$_3$ or by the electric field, inducing the DM interactions between the sublattices. The MFT provides a very reasonable description for the DM interactions in these systems. The crucial point is to take into account the SO coupling of the heavy iodine atoms, which largely contributes to the DM parameters within MFT. Then, the exact technique supplemented with the downfolding of the ligand spins does not change significantly the MFT results. This is in strike contrast with the behavior of isotropic exchange interactions in CrI$_3$, where MFT and the exact technique provide quite a different description.

\par Finally, we have shown how the spin-current model for the DM interactions can be derived starting with the MFT based expression and considering the limit of infinite xc fields. Thus, it can be viewed as the relativistic counterpart of the double exchange mechanism. This also limits the applications of the spin-current model to metallic systems with the strong Hund's rule coupling (resulting in the high-spin state of magnetic ions, similar to the Mn$^{3+}$ ions in manganites~\cite{DE}). On the other hand, the considered in the present work MFT and the exact techniques offer more general framework for the calculations and analysis of the DE interactions.


\begin{thebibliography}{99}
\bibitem{footnote}
Consider the perturbation of the xc field caused by small rotations away from the $z$ axis forming the spin spiral in the $xy$ plane,
$$
\delta \hat{H} =
\left(
\begin{array}{cc}
0 & e^{-\boldsymbol{q} \cdot \boldsymbol{R}_{i}} \\
e^{\boldsymbol{q} \cdot \boldsymbol{R}_{i}} & 0
\end{array}
\right) \theta \hat{b}^{z}.
$$
Then, consider the matrix element of this perturbation between states $| C_{m \boldsymbol{k}}^{1} \rangle = | C_{m \boldsymbol{k}}^{\uparrow} \rangle + | \delta C_{m \boldsymbol{k}}^{\downarrow} \rangle$ and $| C_{m' \boldsymbol{k}'}^{2} \rangle = | C_{m' \boldsymbol{k}'}^{\downarrow} \rangle + | \delta C_{m' \boldsymbol{k}'}^{\downarrow} \rangle$, where $| C_{m \boldsymbol{k}}^{\uparrow} \rangle$ and $| C_{m' \boldsymbol{k}'}^{\downarrow} \rangle$ are the main components and $| \delta C_{m \boldsymbol{k}}^{\downarrow} \rangle$ and $| \delta C_{m' \boldsymbol{k}'}^{\downarrow} \rangle$ are the corrections caused by the off-diagonal elements of the SO coupling. The same perturbation theory with respect to $\delta \hat{H}$ is used in order to derive Eq.~(\ref{eq:gresponse}). Then, we will have $\langle C_{m \boldsymbol{k}}^{\uparrow} | \delta \hat{H} | C_{m' \boldsymbol{k}'}^{\downarrow} \rangle \sim \Delta(\boldsymbol{k}'-\boldsymbol{k}-\boldsymbol{q})$, but $\langle \delta C_{m \boldsymbol{k}}^{\downarrow} | \delta \hat{H} | \delta C_{m' \boldsymbol{k}'}^{\uparrow} \rangle \sim \Delta(\boldsymbol{k}'-\boldsymbol{k}+\boldsymbol{q})$, where $\Delta$ is the Kronecker delta. Thus, the perturbation will couple three Bloch vectors: $\boldsymbol{k}$, $\boldsymbol{k}$$+$$\boldsymbol{q}$, and $\boldsymbol{k}$$-$$\boldsymbol{q}$. Nevertheless, the contribution $\sim \boldsymbol{k}$$+$$\boldsymbol{q}$ is of the second order in the SO coupling and thus irrelevant for the DM interactions.
\end{thebibliography}
\end{document}